\documentclass[10pt, conference]{IEEEtran}

\usepackage{cite}
\usepackage{amsmath,amssymb,amsfonts}
\usepackage{mathtools}
\usepackage{graphicx}
\usepackage{booktabs}
\usepackage{multirow}
\usepackage{textcomp}
\usepackage{xcolor}
\usepackage{url}
\usepackage{balance}

\DeclareMathOperator*{\argmax}{arg\,max}
\newcommand{\bmU}{\mathbf{u}}
\newcommand{\bmn}{\mathbf{n}}
\newcommand{\bmd}{\mathbf{d}}
\newcommand{\bmq}{\mathbf{q}}
\newcommand{\bmw}{\boldsymbol{\omega}}
\newcommand{\calC}{\mathcal{C}}
\newcommand{\calS}{\mathcal{S}}
\newcommand{\calB}{\mathcal{B}}

\IEEEoverridecommandlockouts

\def\BibTeX{{\rm B\kern-.05em{\sc i\kern-.025em b}\kern-.08em
    T\kern-.1667em\lower.7ex\hbox{E}\kern-.125emX}}

\begin{document}

\title{Low-Overhead IMU-Assisted Predictive Beam Management for Multiband LEO Direct-to-Device Links}

\author{%
Abdulrahman~Al-Hababi$^{1}$,
Meysam~Ghanbari$^{2}$,
Mohammad~Taghi~Dabiri$^{3}$,\\
Mazen~O.~Hasna$^{4}$,~\IEEEmembership{Senior Member,~IEEE},
and~Khalid~A.~Qaraqe$^{2}$,~\IEEEmembership{Senior Member,~IEEE}\\[2pt]
$^{1}$Department of Electrical and Computer Engineering, Texas A\&M University, College Station, TX, USA\\
(e-mail: aalhababi@tamu.edu).\\
$^{2}$College of Science and Engineering, Hamad Bin Khalifa University, Doha, Qatar\\
(e-mails: megh89467@hbku.edu.qa; kqaraqe@hbku.edu.qa).\\
$^{3}$LiFi Research and Development Centre (LRDC), Electrical Engineering Division, Department of Engineering,\\
University of Cambridge, Cambridge, UK (e-mail: md2173@cam.ac.uk).\\
$^{4}$Department of Electrical Engineering, Qatar University, Doha, Qatar\\
(e-mail: hasna@qu.edu.qa).
}

\maketitle

\begin{abstract}
Direct-to-device (D2D) connectivity from low-Earth-orbit (LEO) satellites is moving to Ku band, where a handheld terminal must obtain directional gain from several small phased-array panels distributed around the chassis. Beam management then becomes a joint satellite--panel--beam selection problem with hundreds of candidates, and ordinary hand motion can change the best candidate within one decision interval. This paper proposes a low-overhead predictive beam-management scheme for a multiband LEO D2D downlink in which a low-frequency anchor link carries control signalling and fallback traffic, and a Ku-band link carries broadband data. The handset's inertial measurement unit (IMU) reports attitude and angular rate with a known delay; the scheme extrapolates the delayed attitude to the current orientation, scores all satellite--panel--beam candidates analytically from the satellite ephemeris and panel geometry, and trains only a small candidate set built with a satellite/panel diversity rule under a fixed pilot budget. The Ku-band link is activated only when its predicted post-training rate exceeds the anchor rate by a margin. Trace-driven Monte Carlo simulations with a four-panel handset and two visible satellites show that, with six pilots (0.3\% training overhead), the proposed scheme improves mean goodput by 21.2\% and reduces broadband outage by 57.6\% at 90$^\circ$/s relative to an equal-budget delayed-attitude baseline, and operates within 3.8\% of a zero-overhead oracle.
\end{abstract}


\section{Introduction}
Direct-to-device (D2D) connectivity from low-Earth-orbit (LEO) constellations is moving from narrowband messaging towards broadband service for handheld terminals~\cite{Kodheli2021,Tuzi2023,Abdelsadek2023}. Standardisation has followed the same path: 3GPP extended NR non-terrestrial networks (NTN) from sub-7\,GHz bands in Release~17 to bands above 10\,GHz in Release~18, and Release~19 introduced Ku bands for NR NTN~\cite{3GPPRel19}. At Ku-band, however, a quasi-omnidirectional handset antenna cannot close the link budget. The terminal needs a directional gain on the order of 10\,dBi, and in a handset this gain has to come from several small phased-array panels distributed around the chassis, so that coverage survives arbitrary device orientation and hand grip~\cite{Rezaie2022multipanel,3GPP38901}.

This architecture turns beam management into a joint satellite--panel--beam selection problem. 
The search space grows with the number of visible satellites, handset panels, and beams per panel. In the four-panel terminal considered here, two visible satellites produce 392 satellite–panel–beam candidates, so an exhaustive sweep with 50\,$\mu$s pilots consumes almost 20\% of a 100\,ms decision interval.
Hierarchical search and local tracking reduce this cost~\cite{Giordani2019,Heng2021}, but both assume that the best beam changes slowly relative to the training period. The assumption fails for handheld terminals. Ordinary wrist and arm motion rotates the device at tens of degrees per second or more, moves the satellite direction across panel boundaries, and changes which panel is shadowed by the hand or head~\cite{Fernandes2022,BinIqbal2023}. The LEO D2D channel also differs structurally from terrestrial millimetre-wave (mmWave) links. It is dominated by a single line-of-sight path whose global direction is known from the satellite ephemeris, so the residual uncertainty lies almost entirely in the handset attitude. Commercial handsets already carry an inertial measurement unit (IMU) that measures this quantity directly.

\subsection{Related Work}
\textbf{Sensor-aided beam management.} A large body of work reduces in-band training by exploiting out-of-band or context information, including sub-6\,GHz spatial information~\cite{Ali2018}, radar~\cite{Liu2020}, and, more broadly, integrated sensing and communication~\cite{GonzalezPrelcic2024}. Inertial and orientation information has been used in several ways. Brambilla \emph{et al.}~\cite{Brambilla2019} predicted vehicle pitch dynamics from on-array inertial sensors to steer vehicle-to-vehicle mmWave beams and shorten beam alignment. Ali \emph{et al.}~\cite{Ali2021} fused IMU orientation with reference-signal received power through a particle filter. Rezaie \emph{et al.}~\cite{Rezaie2020,Rezaie2022} used location and orientation to learn a short candidate list for beam selection, and later extended this idea to multi-panel devices, where exhaustive search over every beam of every panel is prohibitive~\cite{Rezaie2022multipanel}. Sensor-based reduction of the UE measurement set has also been patented for terrestrial mmWave multi-panel handsets~\cite{SamsungSensorBM}, using the current orientation rather than a delay-compensated prediction, and without multiple visible satellites or a fallback band. The mobility of UE-side beamforming with multi-panel handsets has also been analysed in the 5G-Advanced context~\cite{BinIqbal2023}. These studies target terrestrial mmWave links with rich multipath, and the learning-based methods require site-specific training data. None of them considers satellite links, multiple visible satellites, or a low-band fallback.

\textbf{Satellite user terminals.}
Electronically steered LEO user terminals are known to combine satellite ephemeris, received-signal-strength perturbation tracking, and gyroscope measurements to estimate and track array orientation~\cite{SpaceXPatent}. Such designs assume a single planar aperture on a fixed or vehicular platform with slow attitude dynamics.
They do not address a rapidly rotating handset with several panels and time-varying hand blockage. More broadly, recent studies of highly directional aerial and satellite links have highlighted the strong sensitivity of narrow-beam operation to platform motion and residual beam misalignment~\cite{Erdem2026Terahertz,Dabiri2026QKD}. Work on broadband handheld LEO access has concentrated on satellite-side techniques, such as distributed and cell-free MIMO across satellites~\cite{Abdelsadek2023} and distributed D2D architectures~\cite{Tuzi2023}, and usually models the handset antenna as a fixed-gain element. A recent survey on pointing errors in narrow-beam links identifies alignment in D2D-NTN as an open problem~\cite{Ghanbari2025}.

To the best of our knowledge, no existing work jointly addresses (i) multi-satellite, multi-panel candidate selection under a fixed pilot budget, (ii) handset attitude prediction from delayed IMU data, and (iii) overhead-aware activation of the Ku-band link over a low-band anchor in LEO D2D systems.

\subsection{Contributions}
The main contributions of this paper are as follows.
\begin{itemize}
\item We model a multiband LEO D2D downlink in which a low-frequency anchor carries control signalling and fallback traffic, and a multi-panel handset receives Ku-band broadband data. The model includes a quaternion-based handset attitude, panel element and polarisation responses, panel-dependent Markov hand/head blockage, and an explicit pilot-overhead term in the effective goodput.
\item We propose a diversity-constrained predictive beam-management scheme. It builds an $L$-pilot sounding set with a satellite/panel diversity rule, so that pilots are not concentrated on adjacent codewords of a single panel, and scores all satellite--panel--beam candidates analytically after extrapolating the delayed IMU attitude with the measured angular rate. The radio training cost is $L$ pilots, independent of the codebook size.
\item We introduce an activation rule that enables the Ku-band link only when its predicted post-training rate exceeds the anchor rate by a margin. The terminal therefore does not spend data time on Ku-band opportunities that cannot outperform the fallback link.
\item Using trace-driven Monte Carlo simulation, we show that with six pilots (0.3\% overhead) the proposed scheme improves goodput by 21.2\% and reduces broadband outage by 57.6\% relative to an equal-budget delayed-attitude baseline at 90$^\circ$/s. Because both methods use the same pilot budget and diversity rule, this comparison isolates the contribution of gyro-based prediction.
\end{itemize}

The remainder of the paper is organised as follows. Section~\ref{sec:model} presents the system model, Section~\ref{sec:scheme} describes the proposed beam-management scheme, Section~\ref{sec:results} reports simulation results, and Section~\ref{sec:conclusion} concludes the paper.

\begin{figure}[t]
\centering
\includegraphics[width=\columnwidth]{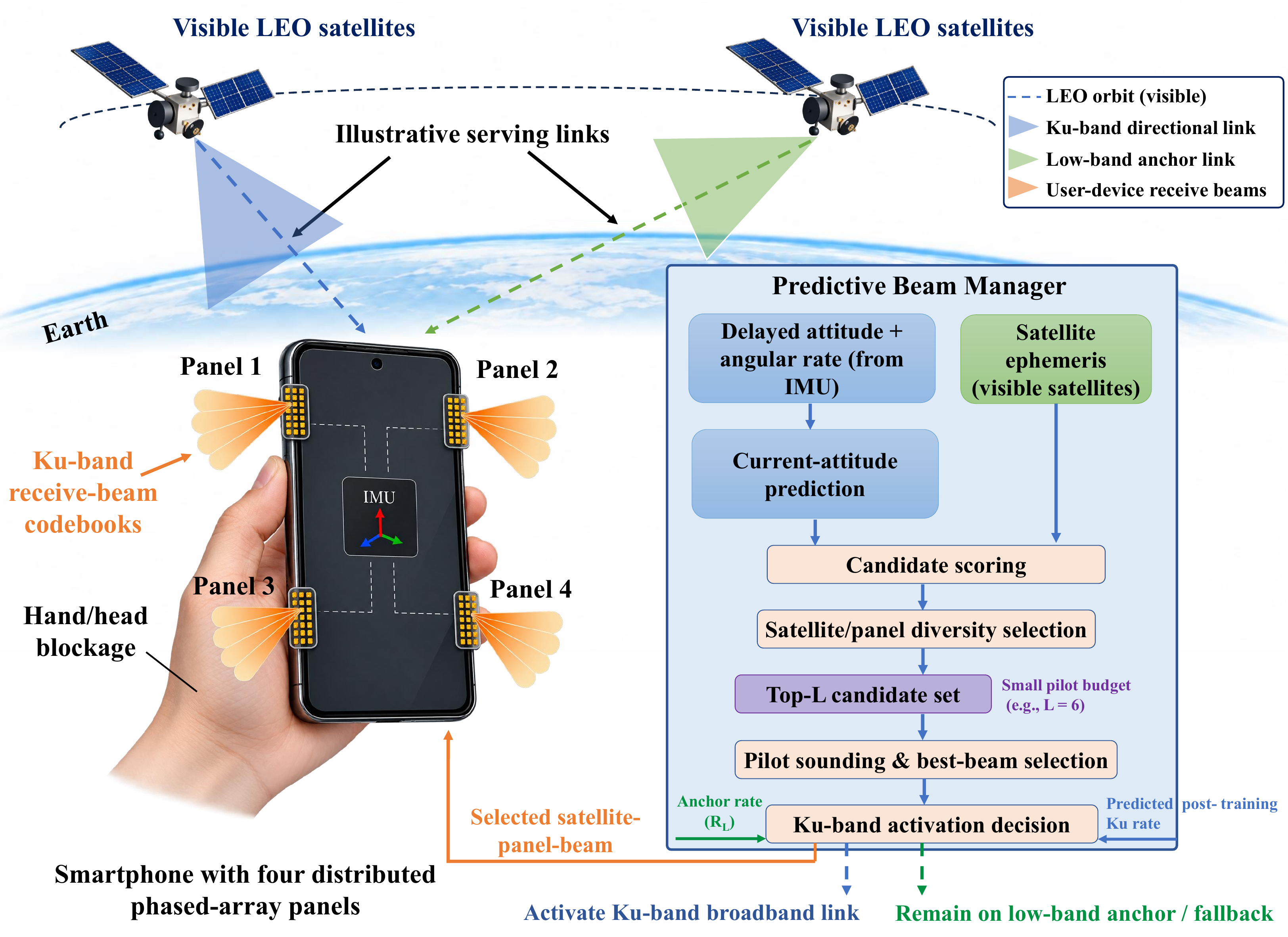}
\caption{Multiband LEO D2D architecture with IMU-assisted predictive beam management for a multi-panel handset.}
\end{figure}

\section{System Model}\label{sec:model}
\subsection{Multiband LEO Direct-to-Device Architecture}
We consider the downlink of a multiband low-Earth-orbit (LEO) direct-to-device system, illustrated in Fig.~1, comprising a set of visible satellites $\mathcal{S}=\{1,\ldots,S\}$ and a handheld user terminal. A robust low-frequency link is maintained as an anchor for control signalling and fallback service, whereas a Ku-band link provides broadband data transmission. The illustrated low-band and Ku-band links represent example serving links only and are not permanently associated with particular satellites; either service may in principle be provided by any visible satellite according to geometry, availability, and the beam-management outcome. The anchor rate is denoted by $R_L$, and the Ku-band bandwidth and carrier frequency are $W_K$ and $f_K$, respectively. Beam-management decisions are taken every $T_d$ seconds.
The handset contains $P$ distributed antenna panels. Panel $p$ has outward normal $\bmn_p$ and a local codebook $\calB_p$. A Ku-band candidate is indexed by
\begin{equation}
c=(s,p,b),\qquad s\in\calS,\; p\in\{1,\dots,P\},\; b\in\calB_p,
\label{eq:cand}
\end{equation}
and the complete candidate set is $\calC$. In the simulated four-panel terminal, each panel contains a $7\times7$ azimuth--elevation grid, giving 196 panel--beam pairs per satellite and 392 candidates for two visible satellites.

Let $\bmU^{g}_s(t)$ be the unit line-of-sight vector towards satellite $s$ in the global frame. The handset attitude is represented by the unit quaternion $\bmq(t)$ and its corresponding device-to-global rotation matrix $\mathbf{R}(\bmq(t))$. The satellite direction in the handset frame is
\begin{equation}
\bmU_s(t)=\mathbf{R}^{T}(\bmq(t))\,\bmU^{g}_s(t).
\label{eq:us}
\end{equation}
For codeword direction $\bmd_{p,b}$, the instantaneous angular mismatch is
\begin{equation}
\epsilon_{s,p,b}(t)=\cos^{-1}\!\big(\bmd_{p,b}^{T}\bmU_s(t)\big).
\label{eq:eps}
\end{equation}
For elevation angle $e_s(t)$, Earth radius $R_E$, and satellite altitude $h_s$, the slant range is
\begin{equation}
r_s(t)=-R_E\sin e_s(t)+\sqrt{(R_E+h_s)^{2}-R_E^{2}\cos^{2}e_s(t)}.
\label{eq:range}
\end{equation}

\subsection{Panel Gain, Blockage, and Link Rate}
The handset gain combines the steered-array pattern, the panel element response, polarisation mismatch, and panel-dependent blockage. Following the usual quadratic approximation to a phased-array main lobe~\cite{Balanis2023}, the beam attenuation is
\begin{equation}
A_{\mathrm{beam}}(\epsilon)=\min\!\Big(A_{\max},\,12\Big(\frac{\epsilon}{\theta_{3\mathrm{dB}}}\Big)^{2}\Big).
\label{eq:abeam}
\end{equation}
For a front-facing panel, the element contribution is
\begin{equation}
G_{\mathrm{el},p}(t)=10\,q_e\log_{10}\!\big(\max\{\bmn_p^{T}\bmU_s(t),\varepsilon_0\}\big),
\label{eq:gel}
\end{equation}
and a fixed front-to-back attenuation is applied when $\bmn_p^{T}\bmU_s(t)\le0$. Consequently, the gain of candidate $c=(s,p,b)$ in dBi is
\begin{equation}
G_c(t)=G_{\max}+G_{\mathrm{el},p}(t)-A_{\mathrm{beam},c}(t)-A_{\mathrm{aux},p}(t).
\label{eq:gc}
\end{equation}
Here, $A_{\mathrm{aux},p}=A_{\mathrm{pol},p}+A_{\mathrm{blk},p}$ combines projected co-polarisation loss and hand/head blockage. Blockage is modelled independently for each panel by a two-state Markov process with stationary blockage probability $p_{\mathrm{blk}}$ and mean blocked-state duration $T_{\mathrm{blk}}$. The blockage loss is drawn once at state entry and remains fixed until the panel returns to the unblocked state.

The free-space loss is
\begin{equation}
L_{\mathrm{FS},s}(t)=20\log_{10}\!\Big(\frac{4\pi r_s(t)f_K}{c_0}\Big),
\label{eq:fspl}
\end{equation}
and the elevation-dependent atmospheric term is represented by
\begin{equation}
A_{\mathrm{atm},s}(t)=\frac{A_{\mathrm{zen}}}{\max\{\sin e_s(t),0.2\}}+A_{\mathrm{rain}},
\label{eq:atm}
\end{equation}
consistent with the elevation scaling used in Earth--space link budgets~\cite{ITUR676,ITUR618}. If $E_s$ is the satellite EIRP, the received power for candidate $c$ is
\begin{equation}
P_c(t)=E_s+G_c(t)-L_{\mathrm{FS},s}(t)-A_{\mathrm{atm},s}(t)-A_{\mathrm{imp}}.
\label{eq:prx}
\end{equation}
The noise power is $N=-174+10\log_{10}W_K+F$\,dBm. With aggregate interference power $I$, the candidate SINR is
\begin{equation}
\gamma_c(t)=\frac{10^{P_c(t)/10}}{10^{N/10}+10^{I/10}}.
\label{eq:sinr}
\end{equation}
The corresponding candidate rate is
\begin{equation}
R_c(t)=W_K\min\{\log_2(1+\gamma_c(t)),\eta_{\max}\}.
\label{eq:rate}
\end{equation}
If a beam manager probes $M_t$ candidates using pilots of duration $T_p$, its training fraction is
\begin{equation}
\rho_t=\min\!\Big(1,\frac{M_tT_p}{T_d}\Big).
\label{eq:rho}
\end{equation}
After pilot-based selection of $c_t$, the effective goodput is
\begin{equation}
R_{\mathrm{eff}}(t)=\max\{R_L,\,(1-\rho_t)R_{c_t}(t)\},
\label{eq:reff}
\end{equation}
where the low-band anchor prevents complete service interruption when the predicted Ku-band link is unreliable.

\section{IMU-Assisted Diversity-Aware Predictive Beam Management}\label{sec:scheme}
The central difficulty is that the best satellite--panel--beam combination can change within an IMU reporting delay. An exhaustive sweep resolves this uncertainty but requires $|\calC|$ pilots per decision. Conversely, selecting neighbouring beams around the previous choice has low overhead but becomes unreliable under rapid handset rotation. The proposed scheme uses delayed attitude and angular-rate measurements to predict the current orientation and restrict training to a small, panel-diverse candidate set.

\subsection{Attitude Prediction and Candidate Scoring}
Let the IMU deliver attitude $\tilde{\bmq}(t-dT_d)$ and body angular velocity $\tilde{\bmw}(t-dT_d)$ with delay $d$ decision intervals. The IMU itself samples at several hundred hertz; $dT_d$ models the age of the attitude estimate at the instant the sounding set is committed, which is set by the decision cadence and any network-side coordination rather than by the sensor. A first-order quaternion prediction gives
\begin{equation}
\hat{\bmq}(t)=\tilde{\bmq}(t-dT_d)\otimes\exp_q\!\big(dT_d\,\tilde{\bmw}(t-dT_d)\big),
\label{eq:qpred}
\end{equation}
where $\otimes$ denotes quaternion multiplication and $\exp_q(\cdot)$ maps a rotation vector to a unit quaternion. Predicted candidate rates $\hat R_c(t)$ are then obtained from \eqref{eq:us}--\eqref{eq:rate}, using $\hat{\bmq}(t)$, statistical panel-blockage loss, and a conservative SINR margin $\Delta\gamma$.

Simply taking the globally strongest $L$ predicted beams often allocates most pilots to adjacent codewords on the same panel. This is undesirable when the dominant panel is blocked or the attitude estimate crosses a panel boundary. We therefore partition $\calC$ into $SP$ satellite--panel groups. The best predicted codeword of group $(s,p)$ is
\begin{equation}
b^{\star}_{s,p}(t)=\argmax_{b\in\calB_p}\hat R_{s,p,b}(t).
\label{eq:groupbest}
\end{equation}
The group winners are ranked by their predicted rates, and one candidate is assigned to each of the strongest groups until either the pilot budget $L$ or all groups are exhausted. If $L>SP$, the remaining positions are filled by the highest-scoring candidates not already selected. The resulting set $\calC_L(t)$ therefore preserves satellite/panel diversity without increasing the pilot budget. Near a panel boundary, an attitude error larger than the angular distance to the boundary moves the true best candidate to a different panel; allocating one pilot per group guarantees that this panel is still sounded, so the miss probability is bounded by the within-panel codeword error alone.

\subsection{Pilot Selection, Activation, and Fallback}
Only candidates in $\calC_L(t)$ are sounded. The selected beam is
\begin{equation}
c^{\star}_t=\argmax_{c\in\calC_L(t)}R_c(t),
\label{eq:select}
\end{equation}
and the training overhead becomes
\begin{equation}
\rho_{\mathrm{pred}}=\frac{LT_p}{T_d},
\label{eq:rhopred}
\end{equation}
which is independent of the full codebook size. Ku-band transmission is activated only if
\begin{equation}
(1-\rho_{\mathrm{pred}})\max_{c\in\calC_L(t)}\hat R_c(t)>\mu R_L,
\label{eq:activate}
\end{equation}
where $\mu>1$ is an activation margin. Otherwise, the terminal remains on the low-band link. This rule avoids spending data time on a predicted Ku-band opportunity that cannot exceed the fallback rate after training.

The online radio-training cost is $L$ pilots rather than $|\calC|$. Candidate scoring still requires $\mathcal{O}(|\calC|)$ lightweight arithmetic operations, but it replaces RF measurements by calculations based on geometry and IMU data. The method is compared with low-band-only operation, reactive local tracking with periodic resets, a two-stage hierarchical sweep, exhaustive sweeping, delayed-attitude selection without gyro extrapolation, and a zero-overhead oracle. The delayed-attitude method uses the same $L$ pilots and diversity rule, so its comparison with the proposed method isolates the benefit of angular-rate prediction.

\section{Simulation Setup and Results}\label{sec:results}
\subsection{Simulation Setup}
The system was evaluated using a trace-driven Monte Carlo simulator in MATLAB. Two LEO satellites at an altitude of 550\,km were considered. The Ku-band carrier and bandwidth were 14\,GHz and 200\,MHz, respectively; satellite EIRP was 83\,dBm, receiver noise figure was 6\,dB, implementation loss was 2\,dB, and aggregate interference-to-noise ratio was $-3$\,dB. The zenith atmospheric and rain losses were 0.8 and 1\,dB. Spectral efficiency was capped at 8\,bit/s/Hz, while the low-band fallback rate and broadband threshold were 8 and 50\,Mbps. The activation margin was $\mu=$\,1.10 and the predictive SINR margin was $\Delta\gamma=$\,2.5\,dB.

The handset used four distributed panels with maximum gain 12\,dBi, 18$^\circ$ half-power beamwidth, and a 15$^\circ$ codebook spacing over $[-45^\circ,45^\circ]$ in both local angular dimensions. The front-to-back and maximum beam attenuations were 35 and 32\,dB. The stationary blockage probability was 0.15, the mean blockage duration was 0.7\,s, and blockage loss was Gaussian with mean 18\,dB and standard deviation 4\,dB, truncated to 8--30\,dB.

The decision period was 100\,ms and each pilot occupied 50\,$\mu$s. Thus, the proposed $L=6$ search used 0.3\% of each decision interval, compared with approximately 6.76\% for hierarchical search and 19.6\% for a 392-candidate sweep. The attitude and angular-rate noise standard deviations were $2^\circ$ and $3^\circ$/s, with one-decision-interval latency. The orientation excursion was fixed at $40^\circ$ RMS for all cases; only its temporal frequency was varied to produce RMS angular speeds of $\{5,15,30,45,60,90\}^\circ$/s. This construction prevents changes in the marginal orientation distribution from being mistaken for a mobility effect. Each point averages 100 independent 20-s trials.

\begin{figure}[t]
\centering
\includegraphics[width=\columnwidth]{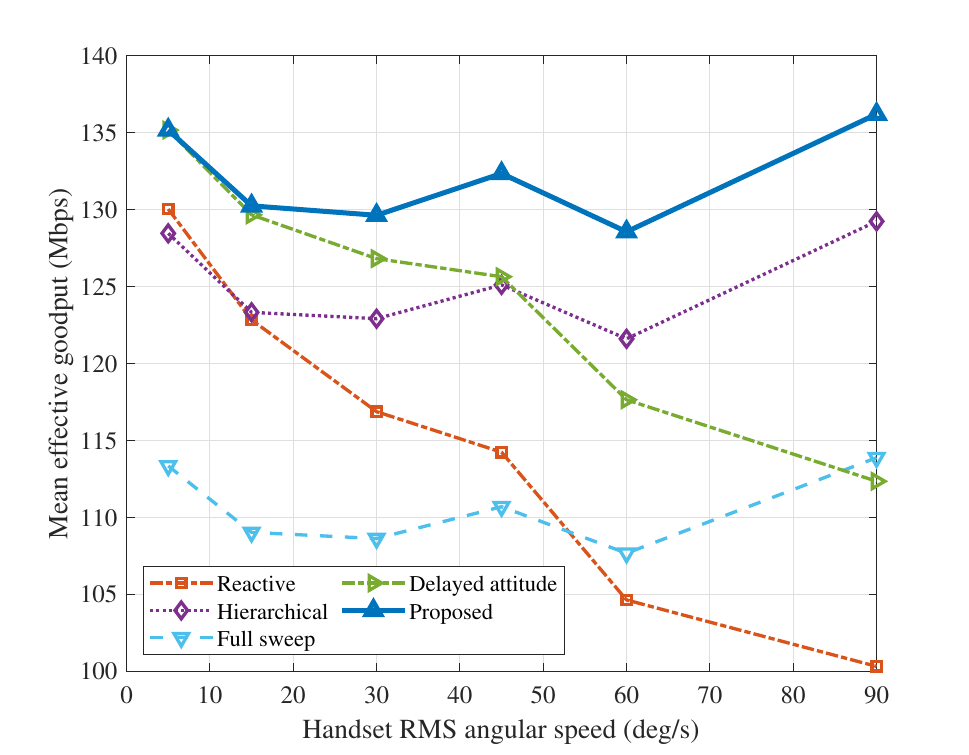}
\caption{Mean effective goodput versus handset RMS angular speed. The proposed method uses $L=6$ predictive pilots.}
\label{fig:goodput}
\end{figure}

\begin{figure}[t]
\centering
\includegraphics[width=\columnwidth]{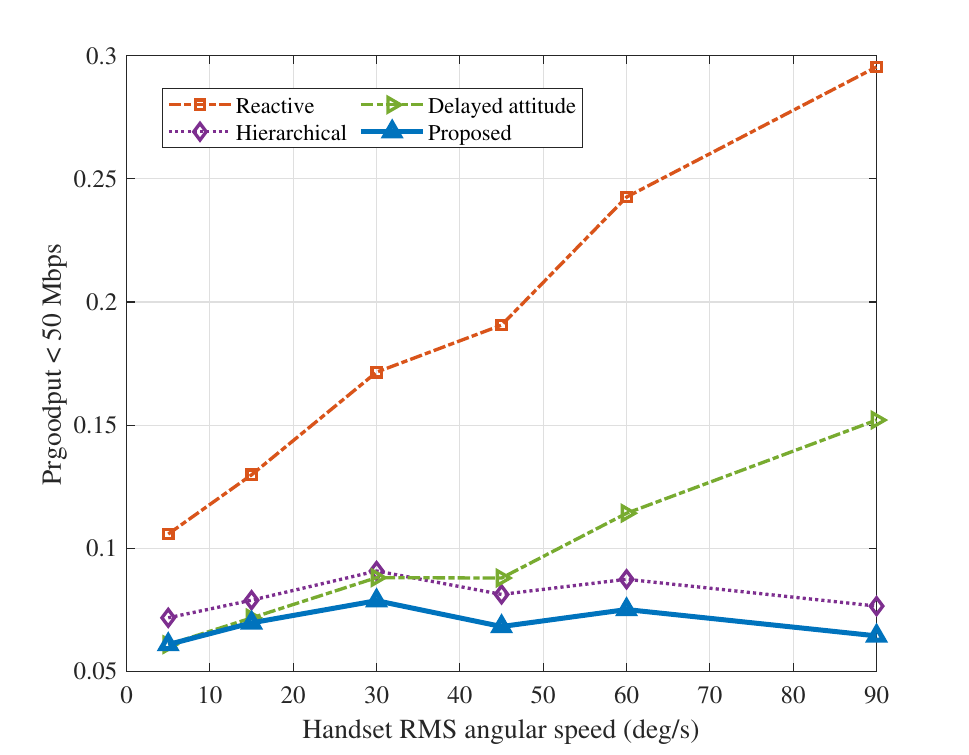}
\caption{Broadband outage probability versus handset RMS angular speed for the four adaptive beam-management methods.}
\label{fig:outage}
\end{figure}

\begin{figure}[t]
\centering
\includegraphics[width=\columnwidth]{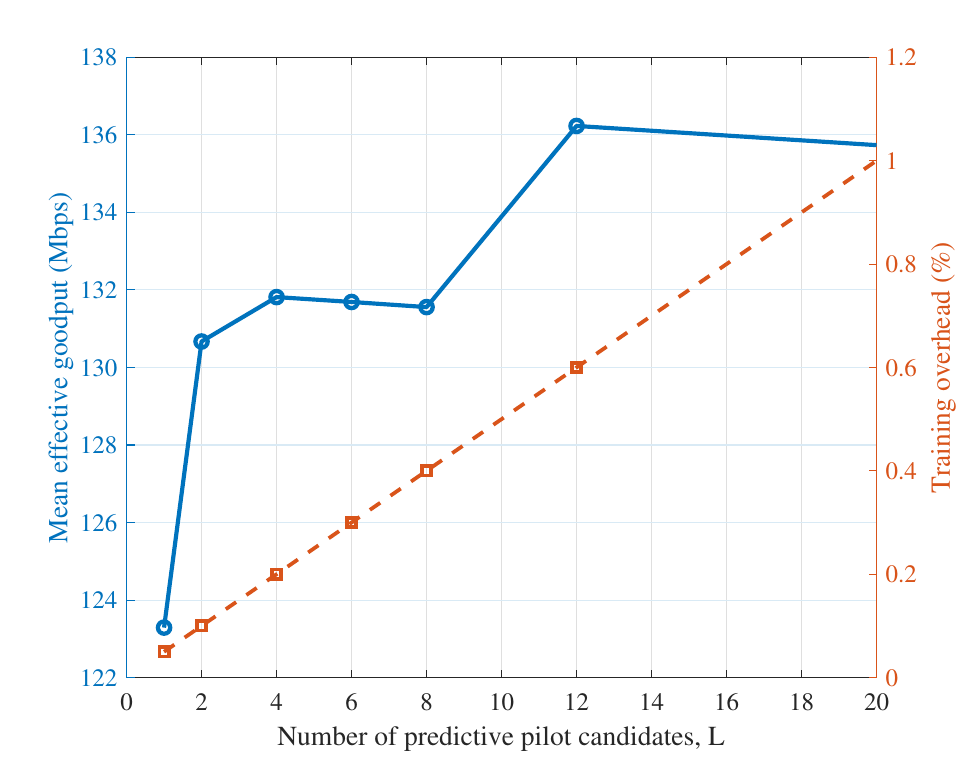}
\caption{Goodput--overhead tradeoff versus the number of predictive pilot candidates at $45^\circ$/s.}
\label{fig:tradeoff}
\end{figure}

\begin{table*}[t]
\centering
\caption{Performance at moderate and high handset angular speeds (mean goodput is reported with the 95\% confidence-interval half-width)}
\label{tab:results}
\begin{tabular}{llcccc}
\toprule
$\omega_{\mathrm{RMS}}$ & Method & Goodput (Mbps) & Outage & Overhead (\%) & Capture (\%)\\
\midrule
\multirow{6}{*}{$45^\circ$/s}
 & Reactive tracking    & $114.24\pm4.93$ & 0.1906 & 2.23  & 68.46\\
 & Hierarchical sweep   & $125.11\pm4.68$ & 0.0813 & 6.76  & 87.50\\
 & Full Ku sweep        & $110.70\pm3.99$ & 0.0840 & 19.60 & 100\\
 & Delayed attitude     & $125.64\pm4.48$ & 0.0879 & 0.30  & 69.21\\
 & Proposed predictive  & $132.35\pm4.81$ & 0.0682 & 0.30  & 81.11\\
 & Zero-overhead oracle & $137.64\pm4.97$ & 0.0579 & 0     & 100\\
\midrule
\multirow{6}{*}{$90^\circ$/s}
 & Reactive tracking    & $100.32\pm4.44$ & 0.2955 & 2.23  & 54.41\\
 & Hierarchical sweep   & $129.24\pm4.88$ & 0.0765 & 6.76  & 88.63\\
 & Full Ku sweep        & $113.88\pm4.20$ & 0.0805 & 19.60 & 100\\
 & Delayed attitude     & $112.35\pm4.07$ & 0.1521 & 0.30  & 49.15\\
 & Proposed predictive  & $136.20\pm5.00$ & 0.0644 & 0.30  & 80.73\\
 & Zero-overhead oracle & $141.61\pm5.22$ & 0.0549 & 0     & 100\\
\bottomrule
\end{tabular}
\end{table*}

\subsection{Goodput and Training Overhead}
Fig.~2 shows that prediction becomes increasingly valuable as handset dynamics increase. At $45^\circ$/s, the proposed scheme achieves 132.35\,Mbps, compared with 125.64\,Mbps for the equal-budget delayed-attitude baseline and 125.11\,Mbps for hierarchical sweeping. At $90^\circ$/s, its goodput reaches 136.20\,Mbps, representing gains of 21.2\% over delayed attitude and 35.8\% over reactive tracking, and comparable goodput to hierarchical sweeping at one twentieth of its pilot overhead. It remains only 3.8\% below the zero-overhead oracle at both 45 and $90^\circ$/s. The comparison with delayed attitude is the most informative: both methods use six pilots and the same diversity rule, so their separation at high angular speed is attributable to gyro-based extrapolation rather than a larger training budget.

The advantage is not produced by excessive beam training. The proposed overhead is 0.3\%, which is 86.5\% lower than reactive tracking, 95.6\% lower than hierarchical sweeping, and 98.5\% lower than exhaustive sweeping. Table~\ref{tab:results} also shows that the proposed candidate-set capture probability remains close to 81\% at 45 and $90^\circ$/s. Hierarchical sweeping captures the instantaneous best beam more often (88.6\% versus 80.7\% at $90^\circ$/s) because it sounds more than twenty times as many candidates; the advantage of the proposed scheme lies in overhead, not in selection accuracy. In contrast, capture for delayed attitude falls from 69.21\% to 49.15\%, explaining its rapid loss of goodput as the handset rotates faster.

\subsection{Broadband Outage}
The broadband outage probability in Fig.~3 is defined as $\Pr\{R_{\mathrm{eff}}<50\,\text{Mbps}\}$. At $90^\circ$/s, the proposed method gives an outage probability of 0.0644, compared with 0.1521 for delayed attitude and 0.2955 for reactive tracking. These correspond to relative reductions of 57.6\% and 78.2\%, respectively. Its outage is also lower than hierarchical sweeping despite using fewer than one twentieth of the hierarchical pilot budget. The result indicates that IMU prediction is useful not only for average goodput but also for retaining broadband service during fast reorientation.

\subsection{Predictive Pilot-Budget Tradeoff}
Fig.~4 examines the predictive candidate budget at $45^\circ$/s. Moving from one to two candidates produces the largest initial goodput improvement because it introduces satellite/panel diversity. Goodput then changes slowly over $L=4$--8, while the training cost increases linearly as $0.05L$\%. A secondary gain appears at $L=12$, but its 0.6\% overhead is twice that of $L=6$. We therefore use $L=6$ as a deliberately low-overhead operating point rather than claiming it as the absolute goodput optimum. Increasing the budget to $L=20$ provides no further gain after accounting for pilot time.

\section{Conclusion and Future Directions}\label{sec:conclusion}
This paper developed a low-overhead multiband beam-management method for Ku-band LEO direct-to-device links. A low-frequency anchor provides fallback connectivity, while delayed handset attitude and angular-rate measurements predict a small, satellite/panel-diverse set of Ku-band candidates. With six pilots, the method incurs only 0.3\% training overhead and preserves approximately 81\% instantaneous-best-beam capture across the tested angular speeds. At $90^\circ$/s, it improves goodput by 21.2\% over an equal-budget delayed-attitude baseline and operates within 3.8\% of the zero-overhead oracle. Future work will replace the analytical panel patterns and blockage process with measured handset data, investigate correlated hand/head blockage and longer IMU latency, and extend the predictor to satellite handover and experimental hardware validation.

\section*{ACKNOWLEDGMENT}
This work was supported by the Qatar Research Development and Innovation Council (QRDI) under Grant No. NPRP14C-0909-210008 and by research funding from Hamad Bin Khalifa University under the Thematic Research Grant Program Cycle 3. M.~T.~Dabiri acknowledges support from a Marie Sk\l{}odowska-Curie Postdoctoral Fellowship. The content is solely the responsibility of the authors and does not necessarily represent the official views of QRDI.

\end{document}